# Understanding the impact of heavy ions and tailoring the optical properties of large-area Monolayer WS$_2$ using Focused Ion Beam

*Fahrettin Sarcan[1,2,*], Nicola J. Fairbairn[3], Panaiot Zotev[4], Toby Severs-Millard[4], Daniel Gillard[4], Xiaochen Wang[5], Ben Conran[5], Michael Heuken[6], Ayse Erol[2], Alexander I. Tartakovskii[4], Thomas F. Krauss[1], Gordon J. Hedley[3], Yue Wang[1,*]*

[1] Department of Physics, School of Physics, Engineering and Technology, University of York, York, YO10 5DD, United Kingdom

[2] Department of Physics, Faculty of Science, Istanbul University, Vezneciler, Istanbul, 34134, Turkey

[3] School of Chemistry, University of Glasgow, Glasgow, G12 8QQ, United Kingdom

[4] Department of Physics and Astronomy, University of Sheffield, Sheffield S3 7RH, United Kingdom

[5] AIXTRON Ltd., Buckingway Business Park, Anderson Rd, Swavesey, Cambridge CB24 4FQ, Great Britain

[6] AIXTRON SE, Dornkaulstraße, 52134 Herzogenrath, Germany

**ABSTRACT:** Focused ion beam (FIB) has been used as an effective tool for precise nanoscale fabrication. It has recently been employed to tailor defect engineering in functional nanomaterials such as two-dimensional transition metal dichalcogenides (TMDCs), providing desirable properties in TMDC-based optoelectronic devices. However, the damage caused by the FIB irradiation and milling process to these delicate atomically thin materials, especially in the extended area, has not yet been elaboratively characterised. Understanding the correlation between lateral ion beam effects and optical properties of 2D TMDCs is crucial in designing and fabricating high-performance optoelectronic devices. In this work, we investigate lateral damage in large-area monolayer $WS_2$ caused by the gallium focused ion beam milling process. Three distinct zones away from the milling location are identified and characterised via steady-state photoluminescence (PL) and Raman spectroscopy. An unexpected bright ring-shaped emission around the milled location has been revealed by time-resolved PL spectroscopy with high spatial resolution. Our finding opens new avenues for tailoring the optical properties of TMDCs by charge and defect engineering via focused ion beam lithography. Furthermore, our study provides evidence that while some localised damage is inevitable, distant destruction can be eliminated by reducing the ion beam current. It paves the way for the use of FIB to create nanostructures in 2D TMDCs, as well as the design and realisation of optoelectrical devices on a wafer scale.

Focused ion beam (FIB) milling is a powerful lithographic technique with nanometre resolution[1], especially when the samples can potentially be damaged by wet chemicals that are required in electron-beam or photo-lithography processes. FIB lithography is often employed as the all-dry nanofabrication method for plasmonic[2–4], photonic[5,6] and microfluidic[7] devices, in applications such as sensors, photodetectors, and light emitters [8–12]. Alongside these applications, more recently, FIB has shown great potential in surface modification and defect engineering in two-dimensional materials, especially in the transition metal dichalcogenides (TMDCs) family, with the aim to tailor their optical and electrical properties[12–19]. It has been shown that direct ion beam irradiation can controllably create chalcogenide and transition metal vacancies, thus manipulating the material's stoichiometry. Ma *et. al.* showed that argon ion irradiation can modify the optical properties of $WS_2$ monolayers (MLs). Sulphur vacancies in ML $WS_2$ create mid-bandgap states, resulting in enhanced saturable absorption in the near-IR wavelength region[15]. Furthermore, controlling the stoichiometry-based defects can also tune the conductivity of 2D materials, sometimes even resulting in a change from a metallic to insulating phase. Stanford *et. al.* showed that controlling the ion irradiation dose can selectively introduce precise local defects in few-layer $WSe_2$, thereby locally tuning the resistivity and transport properties[20]. Also, focused helium ion beams have been used to deterministically create defects in $MoS_2$, $WS_2$ and $WSe_2$[21]. With helium ion irradiation, other than the expected local defects, the sputtered chalcogenide atoms also create an extended percolating network of defects, which results in edge states throughout the monolayers. Using FIB to control defect generation in 2D materials paves the way to fabricate nanoscale novel optoelectrical devices, including single-photon emitters, memory devices, resistors, and atomically thin circuits[9,20–22].

While FIB is the ideal tool for high-precision nanomaterial manipulation, the main challenge is to understand the potential destruction to the materials, not only locally but also in the extended area, and discover how to minimise undesirable damage. The exceptionally delicate nature of these atomically thin 2D materials means that the lateral impact of sputtered ions and milled (and potentially redeposited) atoms becomes critical. Understanding and managing the lateral impact during the inherently destructive FIB process is key to maintaining high-performance nanodevices based on these 2D materials[23,24]. To the best of our knowledge, there are only two studies in the literature regarding quantifying the lateral damage area of ion beam lithography on 2D materials. Both studies are based on chemical-vapor-deposited (CVD) graphene[25,26]. It has been shown that the ions can reach a maximum lateral distance of approximately 30 µm. While Liao *et al.*[25] attributed this effect to the deleterious tail of unfocused ions in the ion beam probe, Thissen *et al.*[26] believe that sputtered gallium ions in the residual gas are the main cause of the long-distance lateral damage. To qualitatively determine the ion sputtering effect around the milled area, there are other studies of graphene that have used Raman spectroscopy to investigate the intensity ratio between the defect-induced Raman feature (D band) and the so-called native graphene feature, G band [27,28]. The D/G ratio of the Raman features decreased exponentially and reached a minimum value at around 15-30 μm away from the area milled by a focused gallium ion beam.

The above studies are all based on graphene. In 2D TMDCs, although the local defects introduced by the ion beam have been studied, the extended lateral impact of the ion beam milling/lithography process, especially the impact on their light-emitting properties remains unknown. In this letter, we performed focused gallium ion beam lithography on a large area monolayer $WS_2$ with a range of ion beam currents. We report the impact of focused gallium ions on monolayer $WS_2$ at different lateral locations and explore the origins of these effects.

Monolayer $WS_2$ was grown on a 2-inch sapphire substrate by AIXTRON Ltd. using a Close Coupled Showerhead metal organic vapour deposition reactor. The uniformity of as-grown ML $WS_2$ across the 2-inch wafer was assessed by photoluminescence and Raman spectroscopy. Centimetre-sized monolayers of $WS_2$ are transferred from the sapphire substrate onto pre-marked 300 nm $SiO_2$-on-Si substrates, using a wet-transfer process with a polymethyl methacrylate (PMMA) layer as a carrier. PMMA was then removed in an acetone bath.

Figure 1b shows the PL spectrum measured from a pristine ML $WS_2$ on $SiO_2$/Si substrate, using a microPL setup[29]. The 532 nm excitation laser beam is focused to a spot of ~1.5 μm in diameter on the sample. The main PL peak energy of pristine ML $WS_2$ is ~2.03 eV, which corresponds to the neutral excitonic transition [16,30–33], presented as $A^0$ in Figure 1. Over the area of interest (within a circular region from the central milled hole, with a radius up to 300 μm), the $A^0$ peak energy varies ±4 meV and the integrated intensities fluctuate by ±5%; this fluctuation is originated from wafer-scale ML CVD growth and the wet-transfer process. Additionally, we observe PL emission at lower energies (two shoulders in Figure 1b, peaked at 1.98 eV and 1.80 eV respectively), corresponding to trions and defect-bound states in ML $WS_2$[30,31]. The energy difference between the $A^0$ peak (neutral excitons) and the $A^T$ peak (trions) is ~34 meV, in agreement with the literature[30]. The trions in as-grown TMDCs are attributed to native defects and TMDC–substrate interaction[34,35]. The lowest energy peak, defect-bound exciton peak $D^S$, is attributed to S-deficiency induced emission, previously discussed in the S-based TMDCs, such as $MoS_2$ and $WS_2$[13–16,30–33,36].

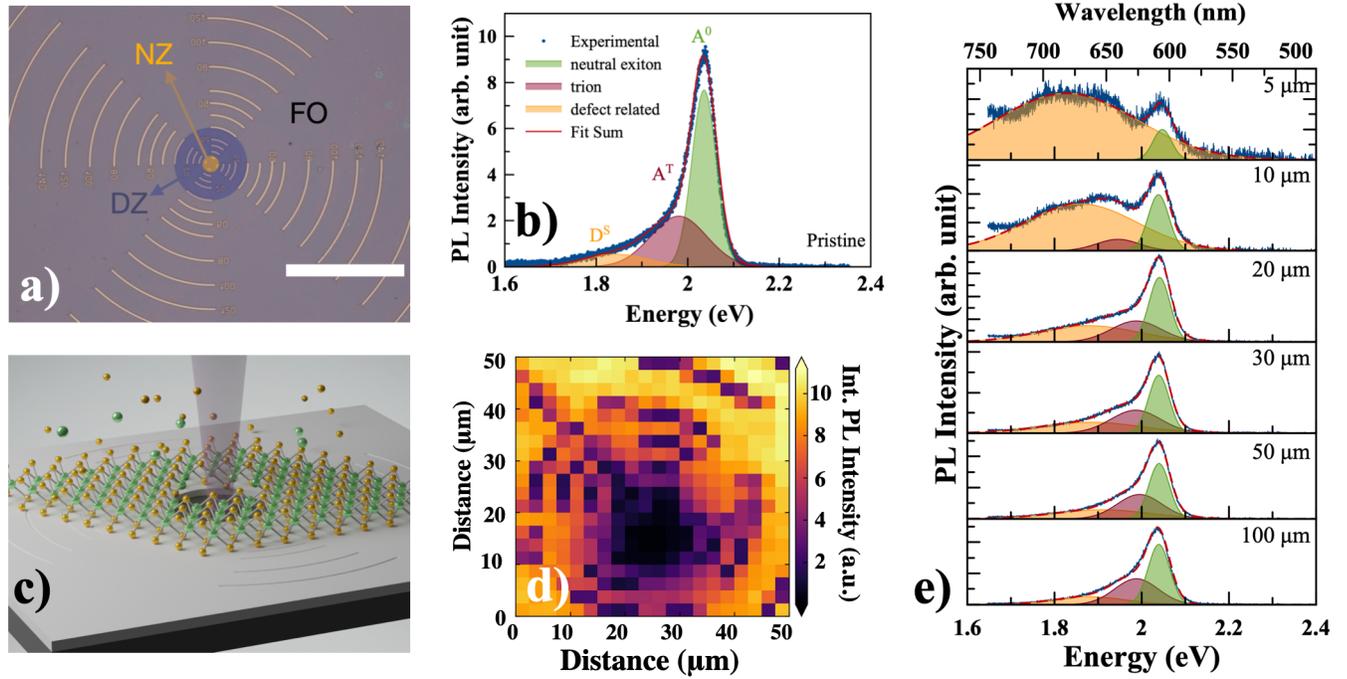

**Figure 1. Steady-state photoluminescence from monolayer WS₂ samples. a)** Optical microscope image of a wet-transferred ML WS₂ on a pre-patterned SiO₂/Si substrate, the scale bar is 100 μm; different colour shades in (a) indicate the three zones - we divide the sample into the near zone (NZ), the dark zone (DZ) and the far-out zone (FO). PL spectra of **b)** as-transferred ML WS₂ (referred to pristine) on a SiO₂/Si substrate, **c)** Schematic drawing of the FIB lithography process on a WS₂/SiO₂/Si sample, d) PL mapping of a FIB-milled WS₂/SiO₂/Si sample (Ga⁺ beam current is 30 pA) and **e)** PL spectra at different locations, with the distance from the FIB milling position labelled on each spectrum. The green, red, and orange colours in (b and e) indicate the three emission peaks originated from neutral exciton ($A^0$), trion ($A^T$) and defect-state ($D^S$), respectively.

A hole with 2 µm in diameter and 50 nm in depth was milled on the WS$_2$/SiO2/Si sample using a dual-beam FIB-SEM system, with a constant dose of 333.4 pC/µm$^2$ (a volume-per-dose of 240 nm$^3$/nC). The lateral impact of the Ga$^+$ FIB process on the optical properties of large area ML WS$_2$ was investigated. Various Ga$^+$ beam currents in the range of 10-3000 pA were tested in the milling process. After the FIB milling process, we remeasured the PL, see Figure 1d and 1e. We divide the lateral area outside the milled hole into three zones: a "near" zone (0-5 µm radius around the milled hole, where 0 is the centre of the hole), a "dark" zone (further 5-30 µm radius away from the hole), and a "far-out" zone (beyond 30 µm from the hole). Figure 1e shows the evolution of the PL spectra moving from the "near" to "far" zone. The total PL intensity and the intensity of the neutral exciton peak (A$^0$) increase. In the "near" zone, the defect emission (D$^S$) dominates. It is obvious that although the milling area is only 1 µm in radius, the laterally affected area is remarkably larger, with a distance up to 30 µm. This extended damage range is caused by the deleterious tail of the ion beam, which consists of unfocused ions and ions in the residual gas in the vacuum system, resulting in the removal of atoms in the ML WS$_2$ and ion contamination on the ML surface.

To investigate the nature of the defect-bound emission D$^S$ at 1.8 eV, we carried out Raman measurements. Figure 2a shows the Raman spectra of pristine and milled ML WS$_2$. Several first and second-order characteristic peaks were observed in the ML WS$_2$ and marked in Figure 2a[32,37,38]. At around 355 cm$^{-1}$, we observed a reduction in the peak intensity and a blue shift of this broad peak. Multi-peak fitting is employed to deconvolve this peak into two sub-peaks, at 349 and 357 cm$^{-1}$ respectively, which correspond to the second-order longitudinal mode (2LA(M)) and in-plane vibrations of tungsten and sulphur atoms ($E^1_{2g}(\Gamma)$) [32,37,38]. The lateral damage in ML WS$_2$ by the FIB milling process is evident by tracking the deconvolved 2LA(M) and $E^1_{2g}(\Gamma)$ peak intensities and widths at various lateral positions (see Figure 2c and 2d). While the peak position of the in-plane vibrations $E^1_{2g}(\Gamma)$ is not affected, the second-order longitudinal mode 2LA(M) is blue-shifted and suppressed around the milled area. This indicates an increased vacancies-phonon scattering within the 30 $\mu$m radius area (in the "near" and "dark" zones), due to the higher density of both sulphur and tungsten vacancies created by the deleterious tail of the ion beam [39].

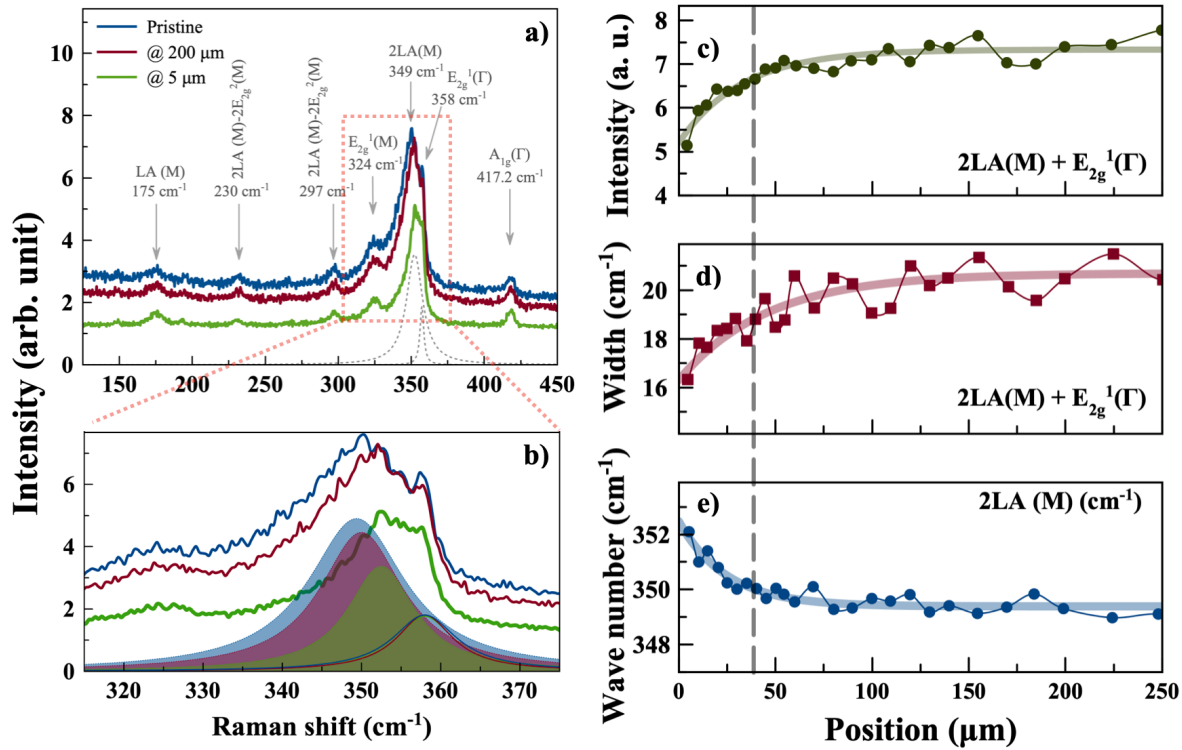

**Figure 2. Raman spectroscopy**. **a)** Raman spectra of pristine and FIB-milled ML WS$_2$ (5 and 200 μm away from the milled area); **b)** Raman peak around 355 cm$^{-1}$ deconvolved it into two peaks: 2LA (M) and $E^1_{2g}(\Gamma)$; **c)** peak intensity, **d)** peak width (full width at half maximum, FWHM) and **e)** peak frequency of 2LA (M) as a function of lateral position.

We further investigated the ion beam damage mechanisms around the milled area by time- and spatially- resolved PL measurements, using a sample-scanning confocal fluorescence time-correlated single-photon counting (TCSPC) setup, shown in Figure 3a. The fluorescence lifetime and decay data are determined using TCSPC with emission detected on single-photon avalanche photodiodes, which are integrated with an inverted confocal microscope with a 100x oil-immersion objective (NA = 1.45). Excitation is provided by a 405 nm laser diode at a 40 MHz repetition rate. As clearly seen in the PL mapping (Figure 3b), there is a ring of higher intensity fluorescence surrounding the milled hole. To help understand the origin of this bright ring in the near zone, we split the collected PL with a dichroic filter centred at 638 nm (being just between the trion and the defect-bound exciton peak wavelengths) to allow us to identify the collected photons at shorter and longer wavelengths simultaneously. With this configuration, we can produce a colour ratio for the PL images, which indicates the "ring" emission is primarily comprised of photons at shorter wavelengths (< 638 nm). Based on the wavelength information, the origin of this "ring" emission could be either neutral excitons or trions, or a combination of both. However, it is known that the trion-induced emission is slower than the neutral exciton emission[40,41]. The time-resolved decay analysis confirms that the "ring" is mainly trion emission. The decay of the ring emission is significantly slower than the emission in the "far-out" zone, which is mainly neutral excitons (see Figure 3d and Table 1). Unlike the native defect and substrate-induced trions, we believe the "ring" originated via two possible routes: first, the positively charged Ga ions depositing and charging the monolayer observed via an electrostatic force microscope, and second, the free carriers accumulating on the edge forming more edge states, resulting in enhanced trion emission; similar effects have been observed in other TMDCs[30,35,41,42].

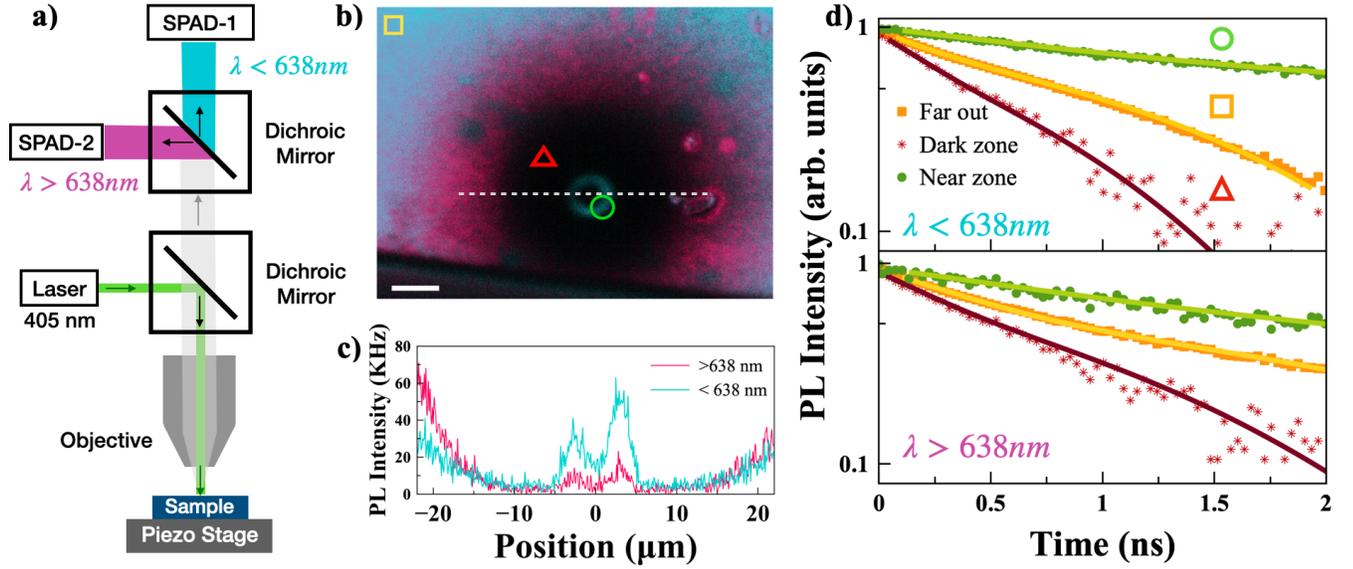

**Figure 3. Time-resolved photoluminescence from FIB-milled ML WS$_2$. a)** Experimental setup of the time-resolved fluorescence confocal microscope; **b)** PL mapping of a FIB-milled sample, with a scale bar of 10 $\mu$m; **c)** PL intensity profile at positions along the dashed line in (b) as a function of lateral position, below (blue) and above (pink) 638 nm; **d)** the time-resolved decay curves at three locations for wavelengths below and above 638 nm: in the far-out zone (yellow squares), in the dark zone (red triangles), and on the "ring" (green circles).

**Table 1**. Exponential lifetimes, τ, and pre-exponential amplitudes (A) of the PL emissions at different locations, fitted with two exponential components

| Wavelength | | Far-out zone, τ | Dark zone, τ | Near zone Ring, τ |
|---|---|---|---|---|
| $\lambda > 638$ nm | A1 | 187 ps (0.79) | 149 ps (0.73) | 552 ps (0.61) |
| | A2 | 905 ps (0.21) | 539 ps (0.27) | 2600 ps (0.39) |
| $\lambda < 638$ nm | A1 | 140 ps (0.77) | 120 ps (0.67) | 586 ps (0.72) |
| | A2 | 429 ps (0.23) | 392 ps (0.33) | 3200 ps (0.28) |

To gain a comprehensive understanding of the lateral damage in the extended area as a function of the ion beam current used, we performed the FIB milling process on ML WS$_2$ samples with a variety of beam currents, from 10 pA to 3 nA. We observed the same trend in the samples milled with beam currents ranging from 10 up to 100 pA, i.e. a linear increase of PL moving away from the milling position in the dark zone, then reaching saturation in the far-out zone, as shown in the normalised PL intensity trend in Figure 4a. While the size of the short-range damage area (the dark zone) remains unaffected by the FIB processes, we observe a current-dependent total emission intensity in the far-out zone. The higher the ion beam current that is used, the less PL intensity is measured. Moreover, the FWHM of the PL spectrum is another indicator of the ion current-dependent damage in the far-out zone. As shown in Figure 4c, the FWHM of the PL spectrum broadens due to the increased defect-bound exciton emissions. It is evident that during the milling process, the ion-beam current plays a crucial role in the lateral damage effect on the ML $WS_2$, contrary to the studies carried out on graphene[25,26]. This secondary far-range damage is due to sputtered $Ga^+$. In a dual-beam-equipped FIB system, the sputtered Ga ions can spread to hundreds of μm away from the milled area. This has been experimentally observed using a time-of-flight secondary ion mass spectroscopy in a similar dual beam FIB[43]. This wider range of ion-sputtering induces secondary damage as the $Ga^+$ and milled atoms that are previously deposited on all the surfaces inside the vacuum chamber (such as the SEM column, the probes etc.) are redeposited on the ML $WS_2$. Redeposition of the $Ga^+$ can be the main reason for the long-range lateral damage, and this can be eliminated by achieving a better vacuum in the chamber. Here, by using a lower beam current, the time required to mill the same amount of material is longer and thereby, it results in additional evacuation of the secondary sputtered ions from the surfaces and the residual gas in the chamber, leading to smaller overall damage of the ML $WS_2$. The $Ga^+$ density on different samples milled with different ion beam currents measured by energy dispersive X-Ray spectroscopy, is noticeably different, as shown in Figure 4d. While only one type of $Ga^+$ distribution (a sharp peak at the milling position) is observed on the low current samples, an additional wide-ranged $Ga^+$ distribution is detected on the high current samples, which causes secondary damage to the TMDC MLs.

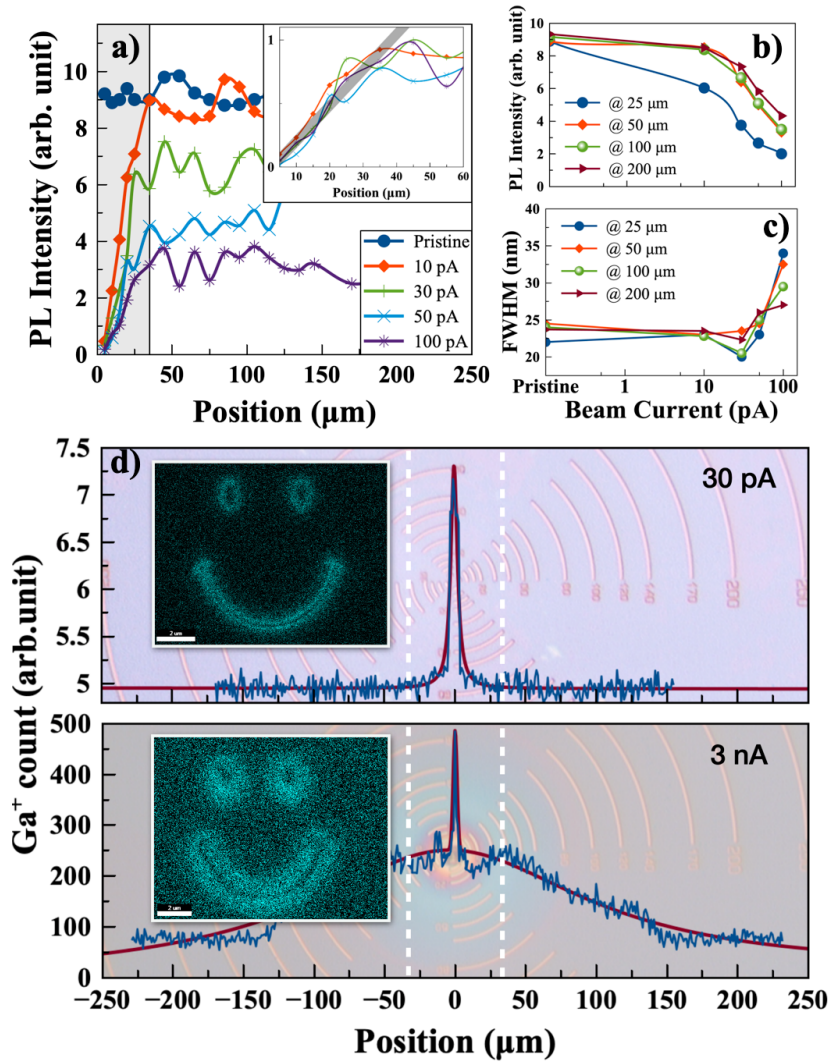

**Figure 4. PL dependency on gallium ion beam current. a)** The emission intensity of the neutral exciton as a function of position, with a range of ion beam current tested; inset is with the intensity normalised to the maximum value on each sample; **b)** Integrated PL intensity and **c)** FWHM of the PL spectra in the "dark" and "far-out" zone as a function of ion beam current. **d)** Distribution of Ga$^+$ on the samples that the centre is milled with 30 pA and 3 nA ion beam current, respectively. We marked the edges of the "dark" zone with two dashed lines. Inset shows two energy dispersive X-Ray (EDX) mapping after a smiley face milled with the corresponding current. The scale bar is 2 µm.

In conclusion, here we have reported the lateral damage effects of a focused ion beam on a large-area CVD grown and transferred monolayer $WS_2$. Three distinct zones situated at different distances from the milling location were clearly identified and the distance-dependent damage mechanisms were explained. The main mechanism for short-range (radius of 0-30 μm) damage is the removal of tungsten and sulphur atoms in the monolayer by the primary ions and ion contaminants on the monolayer surface. The long-range (radius > 30 μm) damage originates from the redeposition of the secondary ions from the residual gas and the ion-electron beam column surfaces. The PL emission was significantly reduced, down to 10% in the short range, however, a bright ring-shaped feature, with half of the original emission intensity, was observed at the edge of the milled area. It is understood that the "ring" is mainly trion emission at a longer wavelength compared to the original emission, due to the charging effect from sputtered $Ga^+$ deposition. This suggests the focused ion beam can be used to tailor the optical properties of TMDC monolayers. We also conclude that while the short-range damage is inevitable, the impact of the weak (secondary) damage can be avoided by reducing the ion beam current during the FIB process. These results pave the way for the use of heavy ion beam-based tools to fabricate novel optoelectrical devices based on the 2D TMDC materials, such as single-photon emitters, memory devices, resistors, and atomically thin circuits, on a wafer scale.

AUTHOR INFORMATION

**Corresponding Author**

* Fahrettin Sarcan: Department of Physics, Faculty of Science, Istanbul University, Vezneciler, 34134, Istanbul, Turkey, orcid.org/0000-0002-8860-4321, email: fahrettin.sarcan@istanbul.edu.tr

* Yue Wang: Department of Physics, University of York, Heslington, York, YO10 5DD, United Kingdom, orcid.org/0000-0002-2482-005X, email: yue.wang@york.ac.uk

**Author Contributions**

X.W., B.C. performed the $WS_2$ wafer growth for this project and the initial characterisation of the wafer, M.H. oversaw the scientific development of 2D material wafer-scale growth at AIXTRON. F.S. and Y.W. fabricated the samples, performed all the SEM-FIB, EDX, EFM measurements, and characterised the steady-state photoluminescence. N.J.F. and G.J.H. performed the time-resolved photoluminescence measurements. P.Z., T.S-M. and D.G. performed the Raman measurements. F.S., Y.W., T.F.K., N.J.F., G.J.H., P.Z., T.S-M., and D.G analysed the results. F.S. A.E., A.I.T., T.F.K., G.J.H. and Y.W. managed various aspects and funded the project. F.S. and Y.W. wrote the manuscript with contributions from all co-authors. Y.W. oversaw the entire project.

**Funding Sources**

Royal Academy of Engineering

Scientific Research Projects Coordination Unit of Istanbul University

The Scientific and Technological Research Council of Turkey

European Research Council

EPSRC

**ACKNOWLEDGMENT**

Y.W. acknowledges a Research Fellowship (TOAST) awarded by the Royal Academy of Engineering. F.S. gratefully acknowledges the support from the Scientific Research Projects Coordination Unit of Istanbul University (FUA-2018-32983) and The Scientific and Technological Research Council of Turkey (TUBITAK) project (121F169). P.Z., T.S-M., D.G. and A.I.T. acknowledge the European Graphene Flagship Project (881603) and EPSRC (EP/S030751/1 and EP/V006975/1). G.J.H. acknowledges EPSRC (EP/V004921/1 and EP/V048805/1). We would like to extend our acknowledgements to a few colleagues at AIXTRON, Clifford McAleese, Sergej Pasko, Simonus Krotkus, for their contribution to WS$_2$ MOCVD growth development.

ABBREVIATIONS

TMDCs, Two-dimensional transition metal dichalcogenides; PL, photoluminescence; FIB, Focused ion beam; ML, monolayer; NZ; near zone, DZ; dark zone, FO; far-out zone, TCSPC; time-correlated single-photon counting, FWHM; full width at half maximum